\documentclass[referee,a4paper,12pt,traditabstract]{swsc} % for a referee version

\usepackage{graphicx}
\usepackage{txfonts}
\usepackage{subfigure}

\begin{document}
   \title{The Planeterrella experiment: from individual initiative to networking}
%   \subtitle{Planeterrella}

   \author{J. Lilensten
          \inst{1}
          \and G. Provan\inst{2} \and S. Grimald \inst{3} \and A. Brekke  \inst{4} \and E. Fl\"uckiger  \inst{5} \and P. Vanlommel \inst{6} \and C. Simon Wedlund \inst{6} \and M. Barth\'el\'emy \inst{1} \and P. Garnier \inst{3}
          }

\institute{Institut de PlanŽtologie et d'Astrophysique de Grenoble,
              38041 Grenoble, France\\
              \email{jean.lilensten@obs.ujf-grenoble.fr}
\and
             Institution University of Leicester, United Kingdom\\
             \email{gp3@ion.le.ac.uk}
\and
             IRAP, 
             \email{Sandrine.Grimald@cesr.fr}
\and
          History of Geoph. and Space Sciences journal,
          \email{asgeir.brekke@uit.no}
\and
          Physikalisches Institut, University of Bern, Switzerland,
          \email{erwin.flueckiger@space.unibe.ch}
\and
          Royal Observatory of Belgium 
          \email{Petra.Vanlommel@oma.be}demonstrated 
\and 
          \email{Cyril.Simon@aeronomie.be}
             }

%%%   \date{Received September 15, 1996; accepted March 16, 1997}

  %\abstract{}{}{}{}{} 
  % 5 {} token are mandatory
 
  \abstract
   {
   Space weather is a relatively new discipline which is generally unknown to the wider public, despite its increasing importance to all of our daily lives. Outreach activities can help in promoting the concept of space weather. In particular the visual beauty and excitement of the aurora make these lights a wonderful inspirational hook.
   \newline
A century ago Norwegian experimental physicist Kristian Birkeland, one of the founding fathers of modern space science, demonstrated with his Terrella experiment the formation of the aurora.  Recently a modernised version of the Terrella has been designed.  This Planeterrella experiment is very flexible, allowing the visualization of many phenomena occurring in our space environment. Although the Planeterrella was originally designed to be small to be demonstrated locally by a scientist, the Planeterrella has proved to be a very successful public outreach experiment.  We believe that its success is due to two main factors  (i) the Planeterrella is not patented and the plans are given freely to any public institution and (ii) the advertisement does not rely on press release, books or web sites but mainly on National and European scientific networks such as COST ES 0803.
      \newline
Today, nine Planeterrellas are operating or under construction in four different countries, and more are in the pipleline. In five years, about 50,000 people in Europe have been able to see live demonstrations of the formation of auroral lights, picture the space environment and get an introduction to space weather with this experiment.  Many more have seen the Planeterrella demonstrated on TV shows.
      \newline
This paper presents the process that led to the making of the Planeterrella and proposes some lessons learned from it.
\newline
(Note: In the following, (jl) refers to the first author of this paper, behind the start of the Planeterrella project.)
  }
     \maketitle


%
%________________________________________________________________

\section{The Terrella}
Kristian Olaf Birkeland (1867-1917) was born in Kristiania, now Oslo where his father was a storekeeper and a ship-owner. He became interested in natural sciences at a very young age and it is claimed that the first thing he bought himself from his own money, was a magnet. Before he was 18 years old he wrote an important paper demonstrating his exceptional skills in mathematics. He became a student in 1885 and was educated as teacher in natural sciences at the Royal Frederiks University of Kristiania in 1890. Birkeland became more and more fascinated by the physics and as soon as he had completed his final university examination, he found himself involved in a study of ''electrical oscillations in a tread of metal."\\
In 1893 Birkeland received a scholarship that gave him the opportunity to visit foreign universities and research groups for carrying his studies further. First he travelled to Paris and met the well known physicist Henri Poincar\'e (1854-1913) who had made pioneering work in calculating the trajectories of charged particles in a magnetic field. While Birkeland was in Paris he got an interest in the motion of how cathode rays (now electrons) move in a magnetic field and made several experiments around this subject.\\
From Paris he went to Gen\`eve where he met professor Lousien de la Rive (1834-1924) the son of the late professor Auguste -Arthur de la Rive (1801-1873) who had studied electric discharges in rarefied gases that had led him to form his theory about the aurora borealis as a discharge phenomenon. Lousien de la Rive, however, together with \'Edouard Sarasin (1843-1917) introduced Birkeland to the electromagnetic waves discovered by Heinrich Hertz (1857-1894) in 1888.\\

   \begin{figure}
   \centering
   \includegraphics[height=4cm]{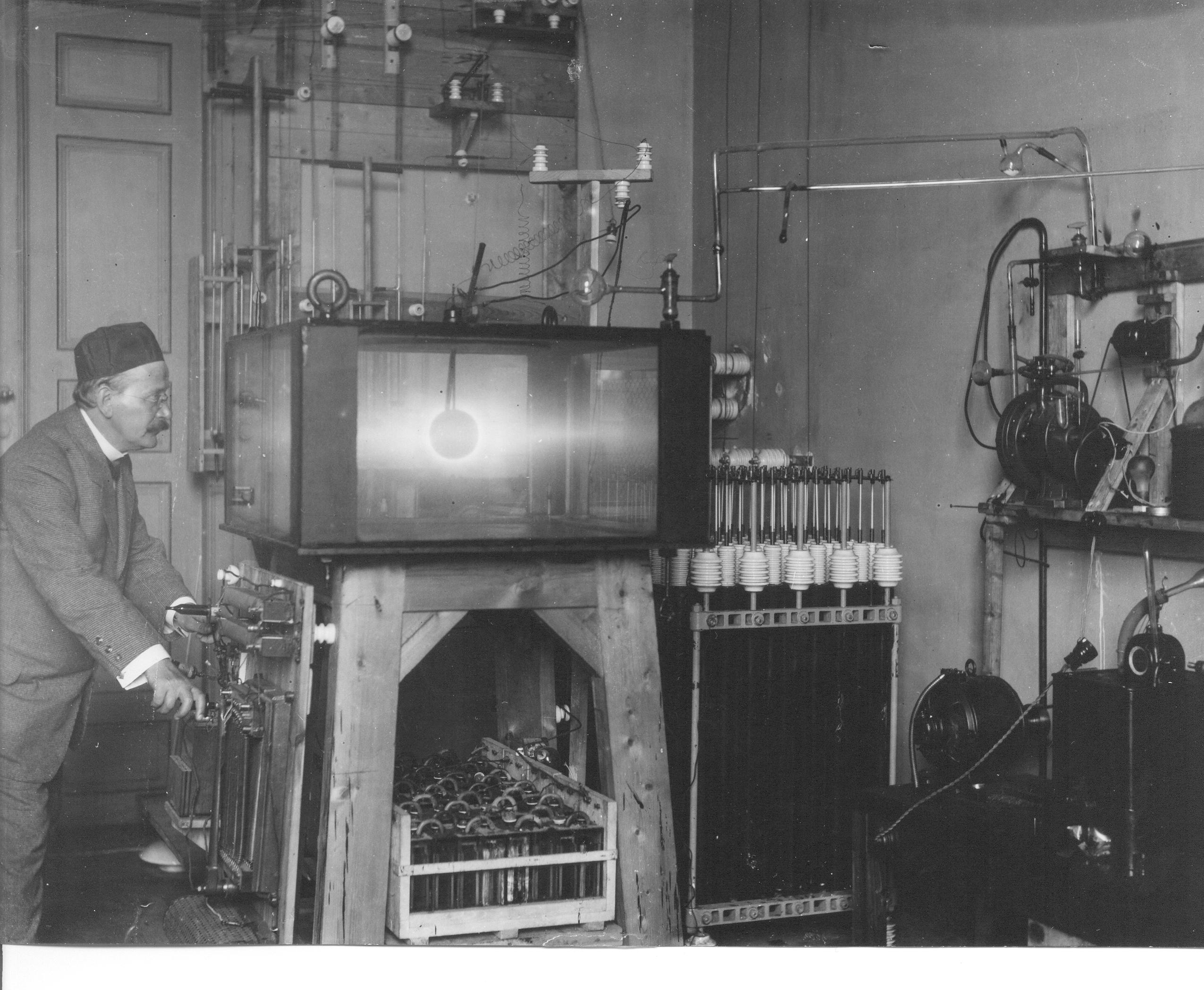}
   \caption{Professor Kristian Birkeland at one of his vacuum chambers performing Terrella experiments.} 
   \label{Terrella_I}
   \end{figure}

When Birkeland came back to Kristiania he carried out several experiments by using different discharge tubes and magnets to observe how cathode rays behaved close to the poles in magnetic fields. He noticed that they were bending and drawn in toward the poles where they created fluorescence illumination. Birkeland published in 1896 his theory about the aurora where he postulated that it was caused by cathode rays streaming out from the Sun and caught by the Earth\textquoteright s magnetic field that forced them to converge toward the magnetic poles where they eventually hit the EarthÕs atmosphere and collided with molecules and atoms forming a fluorescent ring around each pole, aurora borealis and aurora australis (Figure \ref{Terrella_I}).\\

   \begin{figure}
   \centering
   \includegraphics[height=7cm]{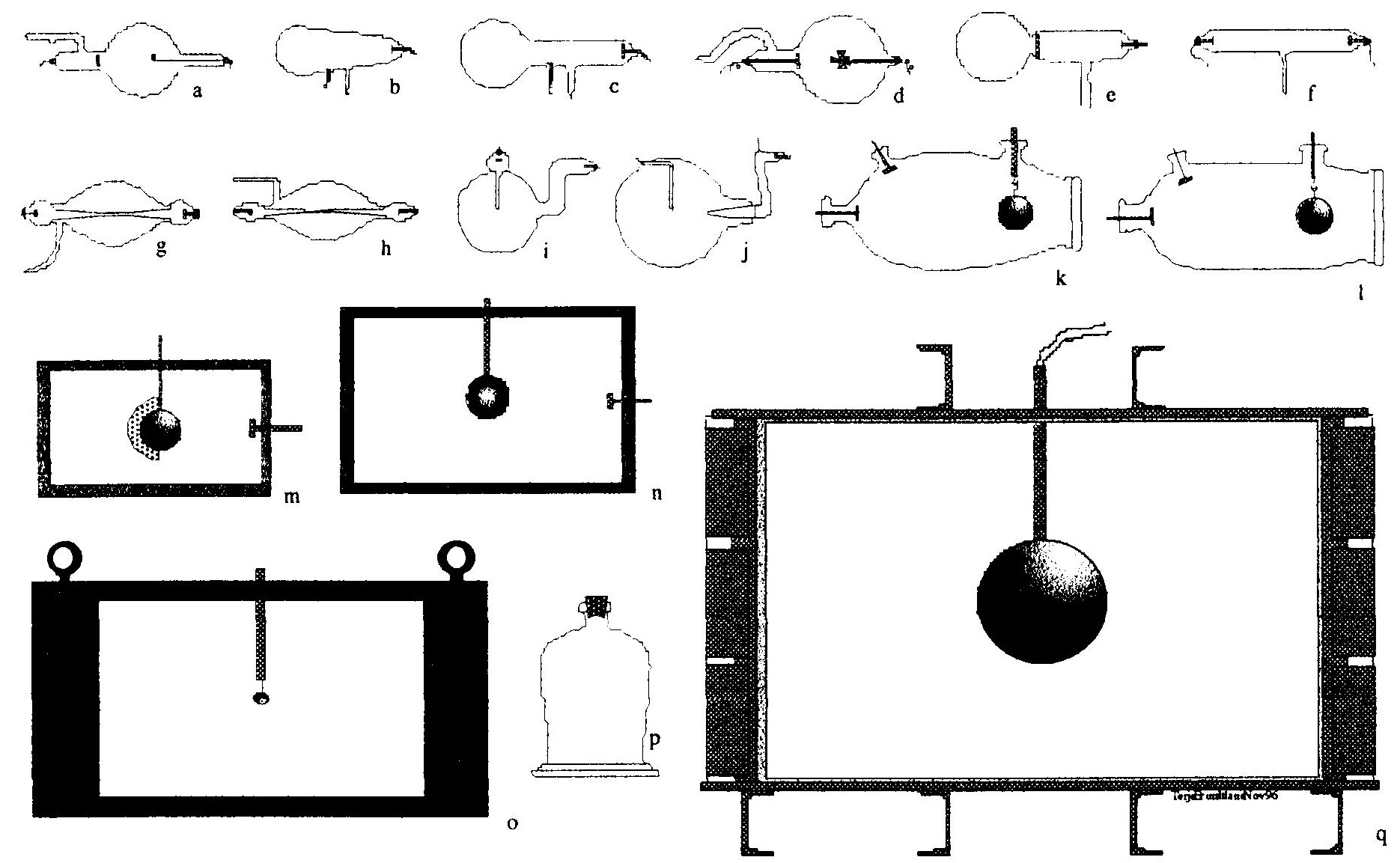}
   \caption{The 17 vacuum chambers used by professor Birkeland for his experiments related to the aurora borealis and other cosmic phenomena.} 
   \label{Terrella_II}
   \end{figure}
   
In response to criticisms of his auroral theory Birkeland in 1901 developed a new set of laboratory experiments using magnetized spheres to simulate cosmic phenomena. At first he named the device a spherical electromagnet, but later on he called it Terrella or ''little Earth". Between 1901 and 1906 he constructed cylindrical and spherical vacuum chambers. Probably from 1909 he designed box-shaped chambers with straight glass walls, the smallest was 22 litres in volume while the largest was referred to as the ''1000 litre prismatic chamber" or "the Universe". Altogether Birkeland used 17 different vacuum chambers (Figure \ref{Terrella_II}). On the technical aspect: the vacuum was about 10 Pa. The tension was larger than 300 V for an intensity on the order of 0.1 to 1.0 mA. The intensity of the magnetic field was approximately 0.5 T at 0.5 cm of the surface of the magnets. Any of these values, easy to reach today, were a challenge at the turn of the 19$^{th}$ century.

\section{The Planeterrella}

\subsection{Rebuilding the Terrella}
In 1996, engineer Terje Brundtland in Troms\o\ (Norway) commenced the restoration of the largest original Terrella built by Birkeland, called ''The Universe" (Brundtland, \cite{Brundtland}). Due to the historical importance of these experiments, the restoration process aimed to re-create the original components whenever possible. Safety, budget and  maintenance constituted additional constraints. The glass was changed to a more modern type, and some of the original wooden supports that were in particularly poor conditions had to be re-build.  The restoration was completed in the late 90Õs and the restored Terrella became a popular exhibit at  the Auroral Observatory in  Troms\o\ . The story of the restoration process may be found at \emph{http://www.mhs.ox.ac.uk/sphaera/index.htm?issue7/articl6}. 
\newline
After visiting the auroral observatory, (jl) built several Terrellas. Amongst these, one is due to the drive of an enthusiastic high school physics teacher together with some of his students. It received a silver medal at the ''Olympics of Physics". Another one was assembled during an international school held at ICTP in Italy (Messerotti et al., 2008).  Based on these Terrellas, (jl) realized that the Terrella experiment could be modernised and improved. This was made with two of the co-authors of this paper (CSW and MB) and in conjunction with a team of collaborators  who are acknowledged at the end of this paper.

\subsection{Planeterrella I}
\begin{figure}
   \centering
   \includegraphics[height=4cm]{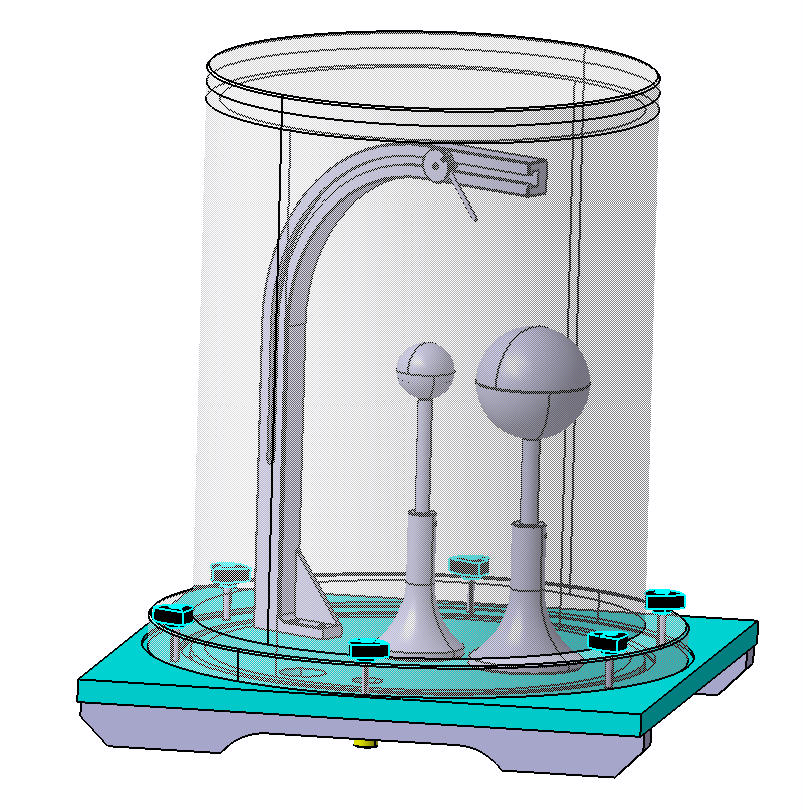}
   \caption{View of the Planeterrella I. The vaccum chamber is in plexiglas.} 
   \label{Planeterrella_I}
   \end{figure}

(jl) considered several ways of improving the Terrella.  Firstly, Birkeland hung the magnetic spheres but such suspension is associated with an inherent fragility as the attachments may break. In particular, suspending the sphere prevents the addition of a second sphere as the two magnetized spheres will attract or repel each other, resulting in the potential snapping of the attachments. This is why Birkeland\textquoteright s successive Terrellas only grew up in size, not in configuration. In the improved design the spheres are placed on supports enabling several spheres to be placed in the chamber simultaneously. (jl) also endeavoured to make the Planeterella as flexible as possible, so that a number of cosmic phenomena can be visualized.  In Birkeland\textquoteright s original experiment, there was an electric duct which exists as well in the Planeterrella in order to re-create Birkeland\textquoteright s observations. However, this duct can now move along a support, and the spheres\textquoteright height may be also changed (Figure \ref{Planeterrella_I}). \\
These modifications thus allowed the visualization of many more space science phenomena than originally observed by Birkeland. It has even been possible to visualize the sunward auroral observations at Uranus and predict the shape of its nightside auroral oval (Lilensten et al. (\cite{lilensten_2009}), Gronoff and Simon Wedlund (\cite{Gronoff_2011})).

\subsection{The goals, the constraints}
Our main goal was to improve Birkeland\textquoteright s experiment in order to provide the best possible auroral demonstration.  It was seen as extremely important to make a tribute to BirkelandÕs original work, and to explain how our work was firmly rooted on his discoveries. Our original target audience were small groups of local people; school classes of all ages, the elderlies, the amateurs in astronomy... Therefore, the target audience size was about 30 people. It was also conditioned by the size of a medium-sized car trunk in which the experiment is transported. Finally, it was important to build an experiment which could be assembled and dismantled without tools. The following Planeterrella experiment was designed within these constraints.\\
The vacuum chamber is 50 liters in volume and 50 cm in diameter. The two aluminium spheres
(Birkeland used copper) have a diameter of 10 and 5 centimeters, respectively.  The Planeterrella is designed to be operated by a professional scientist or engineer at all times. If the Planeterrella is  automatized, the safety constraints would have to be much tighter. The electric power supply is the same as in university student practical work and is secured. The thickness of the vaccum chamber is such that it cannot explode and cannot be broken by people with a hammer. The gases from the vacuum pump are evacuated through an exhaust tube.

\subsection{The cost and the funding}
The main cost is clearly the time spent on designing and building the Planeterrella. Fortunately, (jl)\textquoteright s institute (CNRS) has a laboratory specialized in designing experiments. It charges its manpower about ten times less than a private company. Even so, it costs more than 3000 euros to design the Planeterrella I and an additional  3000 euros for the two following versions. This amount did not include the time of (jl) and two colleagues which amounted to about 200 hours. \\
The experimental equipment costs approximately 8000 Euros. However, it may be possible for much of this equipment to be sourced within many universities, such as the vaccum pump or the power supply, so that the actual cost can be only about 5000 euros. This is a small amount of money compared to scientific experiments, but in our European countries, this is still a large amount of money for outreach activities. \\

Each country has its own funding regulations. In France, if the funding had to be secured in advance through a call and selection by a committee, the Planeterrella would not have been funded. The Planeterrella was then assembled on money from research contracts related to space weather, always reporting afterwards. None of the funding agencies complained about this expenditure. This is not always possible in other countries  where the funding comes through answer to calls. The two ways are complementary and their co-existence allows to take more risks. Now that the experiment has proven to be successful, other institutes find it less difficult to secure funding.

\subsection{Planeterrella II and future prospects}
After two years running Planeterrella I at many different public outreach events and venues, it was decided that a new Planeterrella was desirable.   We wanted the spheres to be movable in order to produce a more dramatic auroral display. The plexiglas vacuum chamber had also become somewhat opaque due to UV emissions,  so it was decided that the new vacuum chamber should be made of glass. Unlike plexiglas, it is possible to make a bell-shaped glass vacuum chamber, allowing a better view from all direction and from above. As will be explained further  below, it was decided to fill the chamber with other gases than air into the vaccum chamber.  A set of reference lines was drawn on the base of the chamber in order to facilitate measurements.

 \begin{figure}
   \centering
   \includegraphics[height=4cm]{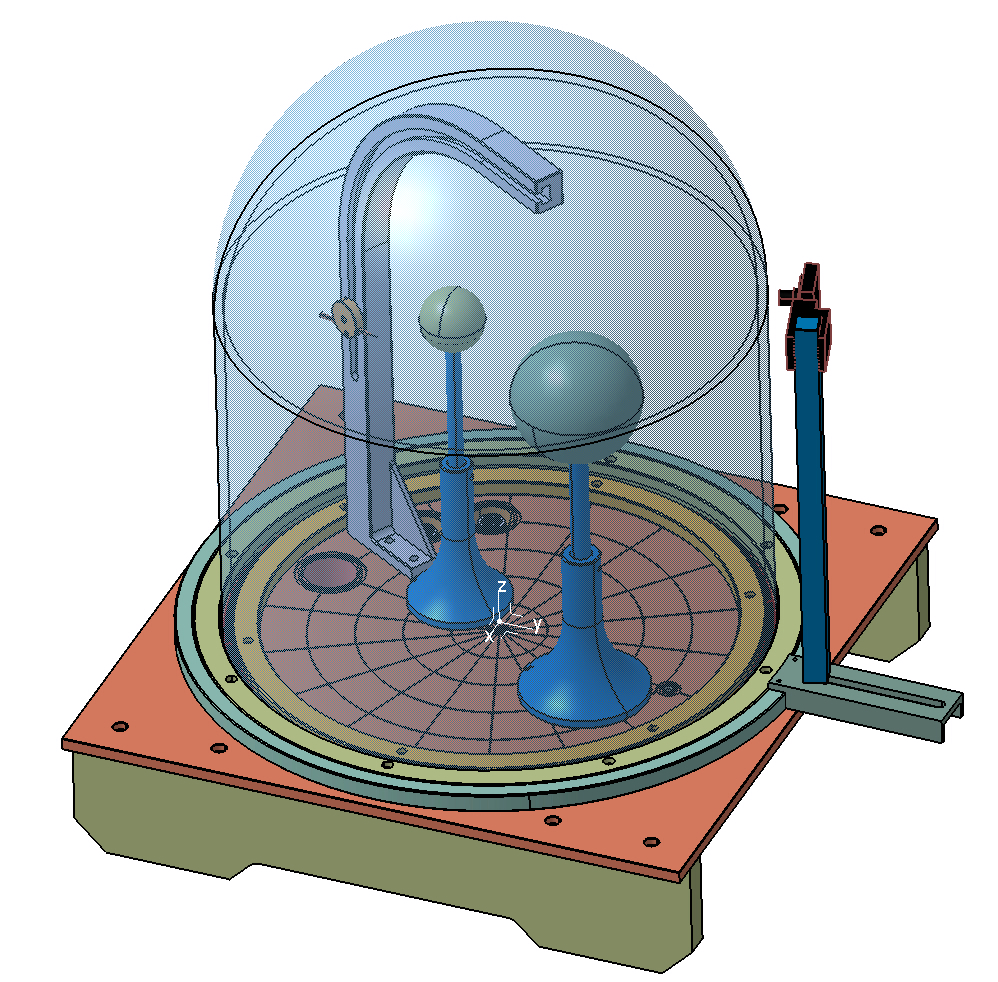}
   \caption{View of the Planeterrella II. The vaccum chamber is in glass.} 
   \label{Planeterrella_II}
   \end{figure}
   
\begin{figure}
   \centering
   \includegraphics[height=4cm]{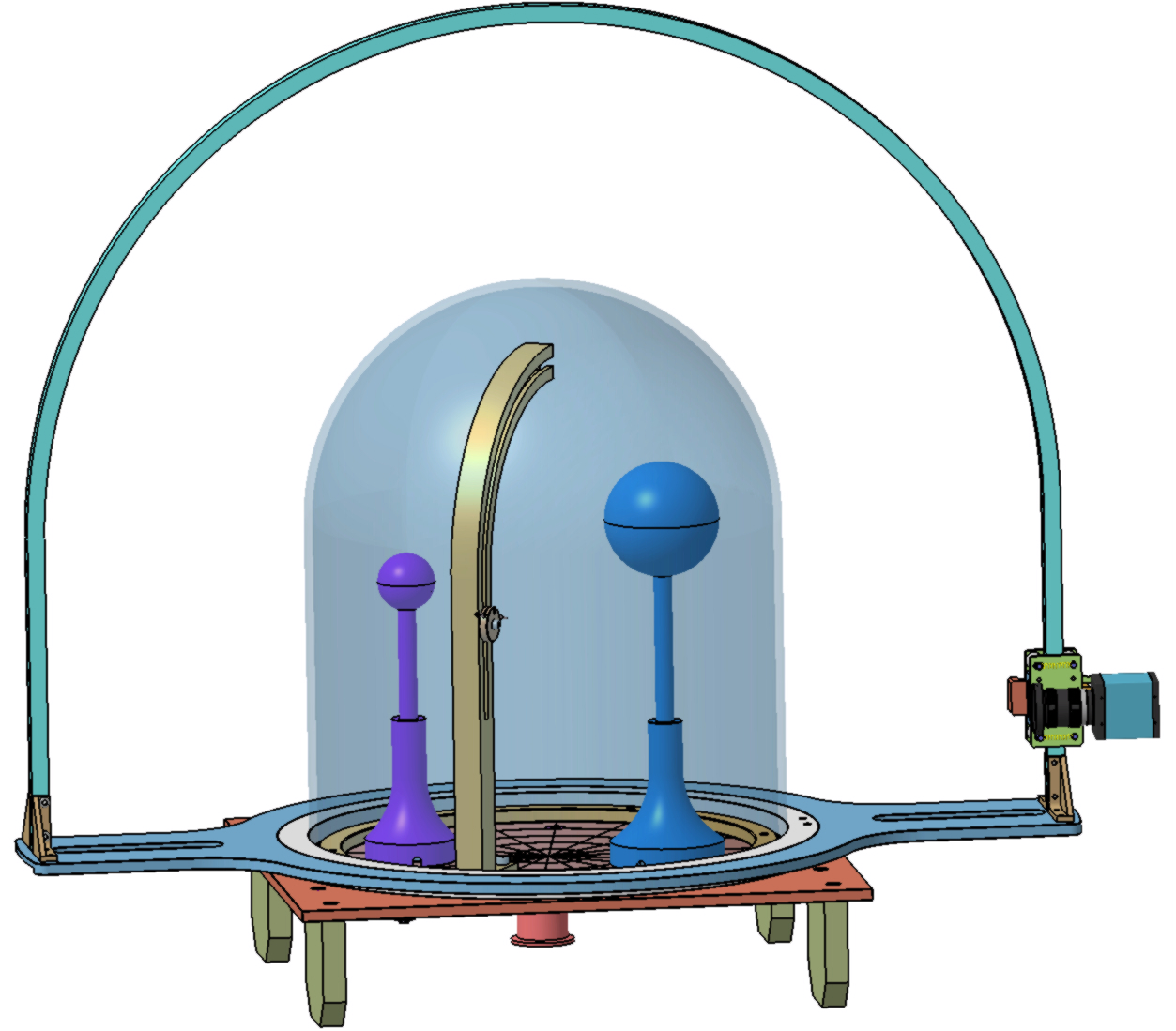}
   \caption{View of the Planeterrella III.} 
   \label{Planeterrella_III}
   \end{figure}

All these improvements were incorporated in Planeterrella II, which was completed in October 2011 (Figure \ref{Planeterrella_II}). Thanks to the movable spheres, it is now possible to demonstrate the existence of the bowshock in front of the sub-solar  magnetopause: it is clearly visible when the spheres are moved towards each other and closely resemble the figures  in many textbooks.\\
A third version of the Planeterrella is currently being designed. Planeterrella III will include
a camera. The main motivation for this is that the Planeterrella has become so popular that it is often shown to more than 100 people at once. Another motivation comes from a new artistic development (see below). In Planeterrella III, the spheres will turn around themselves so that it will be possible to
picture the auroral ovals rotating with the magnetic field or the stellar corona moving with the star (Figure \ref{Planeterrella_III}). In the longer term, improvements will include the use of electromagnets instead of rare
earth ones. The pace of these developments largely depends on funding.

\subsection{Small vs big}
The actual size of the experiment was constrained by the factors described above.  The limitation on the size of the Planeterrella prevents it from being exhibited to too many people at once without a camera and a videoprojector.   However, such an intimate exhibit also has many advantages. The main one is that due to this close proximity the  public actually forget that they look at spheres. As soon as the darkness is established, they often feel as if  they were gigantic persons in space looking at real stars and planets. This works on all kind of people: demonstrations were made in front of groups of  6 to 20 years old scholars with the same effect, to adults, families, people from the country or from cities, to a group of
psychologists and  to scientists. It was shown to the president of the European Space Agency, to the
president of the French Space Agency, to the president of the CNRS. All react the same way, like
the children, with wonder. This is the advantage of the small size.

\subsection{Automatized vs operated}
The main advantage of an automatic experiment is that it can be exhibited in museums. The French space museum (Cit\'e de l'espace in Toulouse) tried to automatize the Planeterrella. This was a failure because automatization results in a sharp increase in health and safety regulations. The spheres must not fall down if someone hits the demonstration table, nobody must be hurt if they try to touch the pump or the tubes etc... Finally, the experiment becomes very static and only a Terrella, not a Planeterrella, can be automatized.  It has been made for the Troms\o\ Museum in Norway by a private company. On the other hand, having an experiment operated by a human has other advantages that are not only linked to the flexibility. First, it allows the visitors to ask questions. But maybe more important and somehow surprising is that it looks very new and modern to the visitors who are all used to numerical demonstrations rather than physical experiments. This is certainly the main lesson learned from this experiment. Finally, it also provides an opportunity to answer the questions on space sciences. These questions may be answered on posters, but no poster can answer in advance all the questions. Moreover, some people can simply not read but are still very curious about science.

   \begin{figure}
   \centering
   \includegraphics[height=4cm]{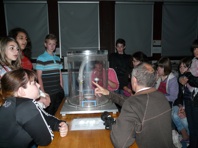}
   \caption{A demonstration with high school students} 
   \label{demo}
   \end{figure}

\subsection{Gentleman agreement vs patent}
When the Planeterrella was created, the CNRS and the University suggested patenting it.
(jl) refused for a variety of reasons. First, Birkeland did not patent the original Terrella. The simple
idea that he could have done so looks ridiculous one century later. Then, the philosophy underlying this new experiment is that public science is costly but the productions of the public science should
remain free for the public. Our work has already been paid by the tax-contributors, and there is no
reason to get a second contribution. This political statement can of course be discussed. Fortunately,  the French administrations (CNRS and University) agreed with (jl) not to patent the Planeterrella.  Instead,  CNRS prepared a ''Gentleman Agreement" to be signed by two parties. On the CNRS side, (jl) provides the plans for free and all the necessary help needed to build a working Planeterrella. The other party agrees not to give the plans to others, to share any improvement and to acknowledge the author of the experiment and his laboratory when exhibiting the Planeterrella.   Scientists are used to networking and sharing ideas, and informal collaborations has been vitally important in disseminating information about the Planeterrella.  \\
A private company has to pay the institute to get the same services.

\section{Networking: sharing and mitigation}
In order to expose the Planeterrella to a wider audience it is important to advertise it. We, as researchers, are not very good at this and our European countries are not good at all in advertising our work compared to the wonderful efforts made in the US. The Planeterrella web site  \emph{(http://planeterrella.obs.ujf-grenoble.fr)} was created with a small budget. The most powerful tool however, which allows European initiatives to survive and spread, is networking. Through a French solar-terrestrial scientific network (PNST), the experiment was shown to the French community. This resulted in 3 copies already existing and one in preparation. In Europe, COST ES 0803 and the Europlanet FP7 action have been the best ground to share it. This is how copies now exist in the UK, Switzerland, Belgium. Agreements have been signed with Italy or are foreseen with Spain. NASA in the US as well as the UCLA university heard of it through these networks and signed the agreement as well. 

\subsection{Other Planeterrellas}
Several Planeterrellas exist in different countries. All together, we estimate that they enabled  auroral phenomena to be introduced to about 50,000 people, not including the many viewers that have seen the Planeterrella on TV shows. In France, it was displayed twice in the second National channel (Antenne 2) in the most popular scientific show in front of probably 5 million people. The second broadcast was for the Christmas special edition. In Germany, it was shown in a movie on Birkeland on the equivalent channel (ZDF). We give here three examples of other Planeterrellas, in UK, Belgium and Switzerland.
\subsubsection{The Leicester example} \label{uk}
In 2010/2011 Dr Gabby Provan and Dr Alan Stocker from the University of Leicester won a Science and Technology Facilities Council (STFC) small award to purchase and display a Planeterrella. The Planeterrella is the only one of its kind in the UK and so allows British visitors to view a unique experiment. Our original aim was to engage and educate a total of about 1000 visitors in the wonders of solar-terrestrial and space physics by using the visual beauty and excitement of the aurora as observed on Earth and other planets. We suggested that the Planeterrella would form the centrepiece of a number of ''Auroral Shows". These shows would include a demonstration of the Planeterrella, a presentation of the latest STFC-supported auroral research and a Questions and Answers session. The "Auroral Shows" would be staged at local schools and colleges, at the University of Leicester (e.g. during open days and school visits), and in other community settings. It was anticipated that ten shows would be held during the course of the 12 month period of the project, the majority of these to be directed at school pupils in the age range 11-18. \\
In order to facilitate the teaching of the Planeterrella within the British school system, additional support was requested and received from the research council which allowed to employ qualified teachers to produce lesson packs which link the Planeterrella to the British National Curriculum.  Four lesson packs have been produced, providing all the material a scientist or engineer needs for a one-hour presentation of the Planeterrella to Key Stages 2,3,4 and 5 (ages 7 to 18) . The lessons ''stand alone" and each pack contains a lesson plan as well as additional materials such as magnets, ''mini-magnetospheres" (a magnet with iron filings suspended in silicon oil) and hand outs etc. The packs also include a teachers\textquoteright guide that can be sent to the school before the visit. \\
All in all the Planeterrella has proved to be a very successful public outreach project and has been exhibited to about 20000 visitors. The space lesson packs have been taught at five schools and have been very well received.  The Planeterrella has been publicised through a host of channels which included writing to individual schools and the Science, Technology, Engineering and Mathematics Network (STEMNET). In addition to the STFC small award, we also won an Institute of Physics University School Link Award which paid for students to assist in demonstrating the Planeterrella. Although the original STFC small award and the Institute of Physics award project periods are now finished, the Planeterrella project is still very much ongoing with almost weekly requests for Auroral Shows. In 2012 the Planeterrella will be exhibited at the British Science Festival in Aberdeen and at the Cheltenham Science Festival, to a combined potential audience of tens of thousands.

\subsubsection{The STCE example} \label{stce}
This is an example where the initial Planeterrella serves as a basis for further improvements. The Solar-Terrestrial Centre of Excellence Ð STCE, a Belgian science and application centre, got the opportunity to build its own Planeterrella experiment.  The STCE provides the financial and human resources. The goals of the STCE Planeterrella are two-fold: science and education. These goals imply requirements. The experiment needs to be mobile, easy to operate, with a flexible and adaptable configuration, safe for the operator and audience, budget friendly.          
With the educative and science goals in mind, a team of scientists, technicians and communication officers was allocated.  In a second step, the concept was rethought and adapted to our goals. The technical plans were drawn and companies contacted. The next step exists of the actual building of the experiment (April 2012). Once the experiment is ready, the real work starts to meet the scientific and educative purposes. \\
We have an up \& fine-tuning of the existing experiment in mind:

\begin{enumerate}
\item Rotating Spheres: Two spheres represent the Sun and the Earth. If we can rotate for example the Earth around its axis (simulation of the daily rotation), one can see that the auroral oval does not rotate with it. 
\item Electromagnets in the spheres: The electromagnets represent the dipolar magnetic field of the Sun and the Earth. Electromagnets allow to change the strength of the magnetic field. If we put a magnetic bar in the coil, we can maybe achieve a broad range in magnetic field strength, going from near 0 Tesla up to 1.5 Tesla. If the electromagnets can rotate, we can simulate a polarity flip similar to the solar magnetic flip every 11 years.  (The capability of changing the tilt between the electromagnet and the rotation axis of the sphere, can contribute to a realistic view of the actual configuration. To simulate active regions andsunspots, we can add a small electromagnet near the surface of the large sphere (i.e. the Sun). This will visualize the coronal loops. 
\item Size of vacuum chamber : We go for a cylindrical chamber with a half a bulb on the top. Due to technical restrictions with the BIRA machinery, the diameter of the basis circle is 46 cm. 
\item Material of the vacuum chamber: We chose for Plexiglas: lighter compared to glass. This increases the mobility of the experiment, less UV transparent, polarizes the light Ð gives the possibility for science about this effect.
\item Simulation of the Dawn-Dusk Electric field:  This field is transversal to the direction Sun-Earth. Extra large electrodes are necessary to create this additional E-field.  A conductive plate as electrode hinders the view. The electrode could be also a conducting grid. 
\item Changeable conductivity of the spheres: A sphere can act as an electron gun. For this, the sphere must be a conductor. The electrons leave the conducting sphere and are guided by the potential/E-field and the magnetic field to the other sphere.  The Earth\textquoteright s ionosphere is a varying conductor: the day-side has a larger conductivity than the night side. Is it possible to change the conductivity of the spheres?
\end{enumerate}

The science part will focus on measurements of emission lines of aurora in different wavelengths. For this purpose, different gases will be injected. The aurora emission lines will be polarized when the bell jar is made of plexiglas. The polarization depends on the magnetic field which is known. With the Planeterrella, we can check if the theory fits the measurements. The experiment allows to check the degradation of filters caused by UV-radiation. This could be of value for space based experiments like PROBA2/SWAP and LYRA. 
The Planeterrella will be used for educative purposes as well. The Planeterrella is an experiment  to communicate about space science in an interactive way between the educator and the audience. We define 5 target audiences. Each target audience needs an specific approach: jargon, language, level.  
\begin{itemize}
\item The non-professional and non-scientific community includes the general public, local institutes, schools and press. These users have to be addressed in their own language with a non-scientific and understandable vocabulary. 
\item The non-professional but with a strong scientific interest. They have often a large knowledge.   
\item The group that looks for a science education for professional reasons. We include students both at undergraduate as at PHD level, scientists that are new to the field of space sciences, as well as people involved in the programming and management aspects of a scientific space science project. 
\item Commercial Entities gather people, companies and institutes that are defined as space weather end users.  
\item Political entities like governments, national agencies, national science policy, international political structures are another important target group.  
\end{itemize}

In this communication project, we have to search and contact our target groups. This is why the requirement of the Planeterrella to be mobile  is crucial. However, we could think of other techniques and tactics to reach our targets through social media like youtube. A demonstration could be taped and put available on the internet. The result is not the same as the real thing, but a movie gives a first idea.\\
The education capabilities are evident: visualizing and explaining aurora, measuring UV transmission and comparing it with the UV radiation outside. The Planeterrella represents a beautiful and accessible introduction to space science and space weather.

\subsubsection{The Bern example} \label{bern}
The Planeterrella project in Bern, Switzerland, was a joint venture between a grammar school (Gymnasium Lerbermatt) and the Space Research and Planetary Sciences Division at the University of Bern. It was initiated in 2010 by Mathieu Dufour, a student who was looking for a fascinating subject for his thesis required for the school leaving examination. Under the lead of PD Dr. Ingo Leya, and in close collaboration with the school teachers Dr. Hans Kammer and Mrs. Irma Mgeladze, the University built two copies of Planeterella I. The engineers and the construction team that usually build space experiments were involved in the project. A shared cost plan was established for the items that had to be purchased (e.g. plexiglas vacuum chamber, vacuum pumps). Most of the other parts were manufactured as a learning example by apprentices in the division\textquoteright s mechanical workshop, and progress of the work was documented in a movie. The Planeterrella at the University was a key attraction of the space physics booth during the ÒNight of ScienceÓ in Bern in September 2011, an event intended to bring science and scientists closer to the general public. This event was attended by more than 7000 visitors. The apparatus is now integrated in the collection of the physics departmentÕs demonstration experiments used in lectures and public outreach activity. With the school version of the Planeterrella, Mathieu Dufour participated in the 2012 Swiss national competition of ''Schweizer Jugend forscht". For his presentation about ÒNorthern LightsÓ, he got the ÒMetrohm special awardÓ and an invitation for the 9th Expo-Sciences Europe 2012 (ESE 2012) in Tula, Russia. \\
Thus, the Bern Planeterella project was extremely successful, and it will continue to be of relevance as a key attraction and demonstration experiment for a broad audience. 

\subsection{The key condition for success}
The most important for a successful experiment  is actually not about money. The most important is about people. Two at least are necessary:
\begin{enumerate}
\item A crazy technician. You have one: they exist in any laboratory. The man or woman to whom one goes when an equipment seems definitely broken, the one able to fix a car with a screw, or a space mission with a piece of paper and some glue.
\item A crazy scientist. Someone who loves to have contacts with the public, who never protests for carrying 100 kg of Planeterrella to a school to show it to 100 children whithout even a journalist, the one who likes to spend week-ends for outreach activities and no-one else will ever know.
\end{enumerate}
If one does not have both people, the experiment is doomed to failure or to stay in a file cabinet for ever. Time, energy and money are lost.
\section{A multipurpose experiment}
Many people have seen pictures of auroras in books or in the Internet but few know how they are created. The Planeterrella allows presenting what could be seen by someone observing the Earth (or other planets) from space and can be used as a starting point to explain the auroral phenomenon. It is therefore a nice tool to show to scholars, to astronomical festivals, elderly houses ... However, it became rapidly popular in the Rh\^one-Alpes area, one of the largest regions in France. Soon, students came to work on it or to use it for their reports. Scientific and soon artistic exhibitions asked to get it. None of these were foreseen. None of the other applications had been foreseen either, and they constitute a source of wonder and perplexity. 
\subsection{Pedagogic}
\begin{figure}
   \centering
   \includegraphics[height=4cm]{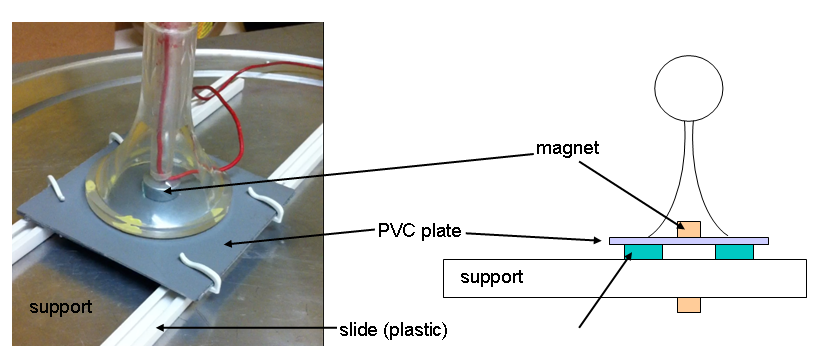}
   \caption{The system developed by the students at the Department of physical measurements in the technological institute of the Toulouse 3 University to make the small sphere to translate along an axis.} 
   \label{toulouse}
   \end{figure}

It rapidly turned out that the Planeterrella can be used as a tool for universities to train future physicists. It  allows presenting and explaining physical processes such as the motion of charged particles in a magnetic field, to visualize astronomical phenomena such as the ring currents circling the planets, the Solar corona or auroral ovals, to study and practice the physics in the vacuum and is also used in different universities to develop spectral analysis. Amongst the different pedagogic uses, we will develop one carried out at the Department of Physical Measurements in the Technological Institute of the Toulouse 3 University in 2010/2012. This institute trains future technicians or engineers. Its sholarship includes a tutored project during the second year which lasts several weeks. The students have to solve a real practical problem, present and defend their work as they shall do it in their future companies. The purpose of the tutored projects is to make the students work in teams, with a large autonomy.\\
Two groups of students dedicated their project to the Planeterrella with two goals: first to prepare its use for future practical courses, and second to improve the experiment. The students had first to understand the experiment: why it was built, which phenomena are shown and how it is constructed. The main pedagogic interest of this tutored project is that the students work to improve an existing experience, which is very rewarding.\\
During a fist period, the students investigated the possible use of this historical experiment for practical courses in vacuum engineering, magnetism, spectroscopy, measurement chains and the physics of electric discharges. In the second part, they studied how it could be improved to show new phenomena and better represent the reality. The second group of students decided to study how to move one of the spheres without opening the chamber and thus without breaking the vacuum. To do so, they had to deal with the magnetic field and the vacuum conditions. They designed a low cost system to allow for the small sphere to easily translate with a reduced friction, thus allowing to observe dynamically various configurations of the plasma interaction. The system consists of two plastic slides usually used for curtains, in which a PVC plate translates. The small sphere is fixed on it (figure \ref{toulouse}). The movement is made using two magnets, one located above the PVC plate and the second one below the Planeterrella support. A translation along a slide is more adapted to practical courses than a free movement inside the chamber to prevent collisions between the spheres or with the chamber wall. The new system is now used in the Toulouse Planeterrella and reveals new phenomena depending on the distance between the spheres.

\subsection{Modeling}
\begin{figure}
   \centering
   \includegraphics[height=6cm]{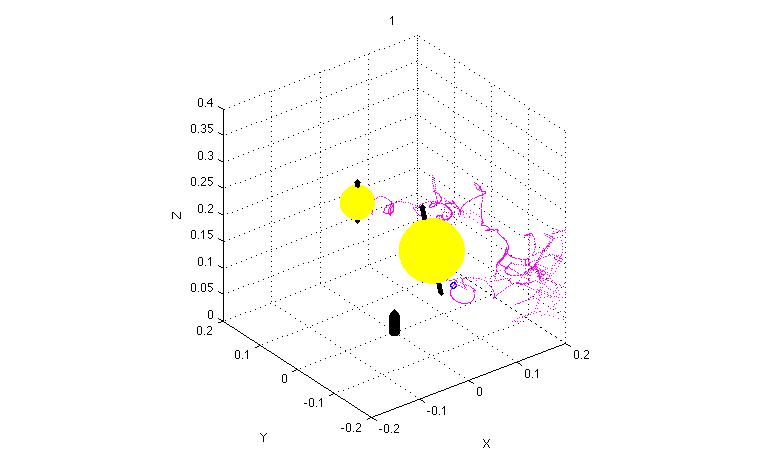}
   \caption{Simulation of the trajectory of a particle with an initial energy of 67eV. In this example, only elastic collisions are taken into account. The two spheres are shown in yellow, with the magnetic axis in black.} 
   \label{simu}
   \end{figure}

In order to make the experiment more understandable for a large public we started to build a numerical model to simulate the trajectories of the particles and to map the regions of emissions.
This model is based on a Monte Carlo simulation technique. The characteristics (region, direction and velocity) of the particle injected in the vacuum are randomly decided. The trajectory is then calculated by solving the movement equation using an adaptative step Runge Kutta 45 method.
Because of the Debye screening effect, we consider that the electric field is fully screened in a very small region close to the cathode. The width of this region has been calculated in Lilensten et al (\cite{lilensten_2009}) and is equal to approximately 3 millimetres. This means that the particles are considered to leave the cathode with a velocity remaining constant after the Debye sphere and are only then affected by the magnetic field. 
This field is considered to be the superposition of the two dipolar fields created by the magnets into the spheres. At this time, the set of cross section taken into account is not complete. Elastic cross sections are considered with an isotropic angle phase function and some inelastic processes are used to obtain a rough evaluation of the emissions. In particular, the secondary electrons are not yet considered.
An example of the computed trajectories is shown in figure \ref{simu}. Depending of the educational level of the public, several phenomena can be illustrated. For postgraduate or Master \textquoteright s degree students, the experiment and this model can be used for a plasma physics practical work. For example to confirm or to infirm the validity of the centre guiding approximation.

\subsection{Artistic}
Many artists are attracted by auroras. Surfing the web, they often happen to reach the Planeterrella site. This is a source of inspiration for paintwork, installations, movies in several countries (France, Belgium, Germany...). The Planeterrella is invited to demos such as the "night of the castles" where it is used as an artwork. Italian sculptors recently claimed that the Planeterrella was not about science but about sculpting the light, which is another way to consider it. The Planeterrella is currently used by a theatre company to prepare a show on Birkeland (in English). In this show under construction, the Planeterrella will be on the stage as one of the actors. For this project, the artists asked to include other gazes in the vaccuum chamber. This was made in Planeterrella II. Art is then now becoming a driver to improve a scientific outreach experiment. 

\subsection{Popularization of space weather, popularization of science}
The Planeterrella allows picturing the Van Allen Belts, the auroral ovals, the Solar corona, the polar cusp, the magnetopause ... All parts of the Earth space environment linked to space weather. It is therefore a great introduction to this discipline. It is very easy thanks to the experiment to demonstrate where and why spacecrafts are exposed. One of the most frequently asked questions is the link between space weather and climate. Although this is highly controversial and no definitive answer can be given, the Planeterrella allows to address this point with a basis to deepen the understanding of the public.\\
Politicians (mayors, deputies aso) constitute a very special public. They must understand things very quickly because they have little time to devote to each concern. The Planeterrella has proven to be quite an efficient tool to inform them in a fast, and accurate manner of what Space Weather is about.
\newline
A report has been published by the UK group (see paragraph \ref{uk}). There was no resources to perform a thorough visitor analysis for all tens of thousands  visitors.  Instead, a focus group of school pupils at one school was interviewed and asked to fill out a questionnaire.  Of the responds:
\begin{itemize}
\item 100\% either strongly agreed or agreed with the statement ''I found the presentation interesting"
\item 100\% either strongly agreed or agreed with the statement ''Science affects everyday life"
\item 75\% agreed with the statement ''I would like to find out more about careers in science"
\end{itemize}

\section{Conclusion}
The Planeterrella finds its roots in Birkeland\textquoteright s work. The five years of demonstrations allowed to draw some lessons:
\begin{itemize}
\item The plans are given freely to any public institute who wants a copy. This is a several tenth of thousands of Euros saving. A Gentleman Agreement has to be signed and there is no patent on it. This is a political choice : Public science is costly but the productions of the public science should  remain free for the public. It explains partly the international success of the Planeterrella. Colleagues are used to share the knowledge. Moreover, they feel secured to get the plans even when no money is yet available. This establishes a climate of confidence. 
\item Science gains at being shared. To that extend, Europe developed several networking tools. This is one of the key factors that makes Europe attractive and dynamics in the different fields of research. Amongst the tools, COST is a very powerful one which helped to develop the Planeterrella in different countries.
\item Most of the actual scientific exhibitions rely on computers, buttons to press in order to get an effect from a robot or a new figure on a screen. This is called ''Interactive Science " and certainly helps attracting children. In this extend, the Planeterrella looks much more like an experiment from the 19$^{th}$ century. However, in the eyes of the visitors, this is very new, totally different from what they are used to and, surprisingly, very modern.
\item One Planeterrella represents a brute investment of 8000 euros. There are now 7 copies completed (more to come) which allowed to show the auroras to about 50,000 people (without mentioning TV shows).  This is a bit less than 0.9 euros per visitor. This is doomed to decrease: in the next future, more Planeterrellas will be built, including one in the largest scientific museum in Paris (le Palais de la d\'ecouverte) which welcomes several tenths of thousands of visitors each year. From this estimate, we consider it worthwhile to invest in outreach in science: small money makes sometimes big effects.
\item To fund the Planeterrella, the post-evaluation has proven to be efficient, at least in the case of the French organization: money comes from other projects, and the evaluators evaluate the use of the money at posteriori. This way is opposite to the call-and-selection process at use in many cases. The two ways do not oppose each other but complete each other. Indeed, if only the second one is at work, the temptation is large to fund only secured projects, i.e. already known things.
\end{itemize}

\begin{acknowledgements}
     Amongst the contributors to the Planeterrella project, (jl) especially thanks Philippe Jeanjacquot, Decybel System as well as Olivier Brissaud, Philippe Chauvin, Karine Briand, Herv\'e Lamy, Laurent Lamy, Baptiste Cecconi, Jean-Francois Donati. Several PhD students were and still are a wonderful support, including Gael Cessateur, H\'el\`ene M\'enager, David Bernard and of course the enthousiastic, imaginative and always supportive Guillaume Gronoff. The Alpes D\'el\'egation CNRS, the Universit\'e Joseph Fourier, the IPAG laboratory and the PNST administrations are also thanked for their constant support.
     The Planeterrella engineer work has been performed at the SERAS / CNRS workshop. This project has been also supported enthusiastically by the COST ES 0803 community. More than 40 students already worked on the project for their studies. Amongst them, we thank particularly Camille Patris, Arthur Birac and Marie Garcon. CSW and PV are indebted to Eddy Equeter and Johan De Keyser (BIRA-IASB) for their unfailing involvement, work, and ideas in the current implementation of the Belgian Planeterrella. \\
     The Planeterrella received the first international price for outreach activities from the Europlanet Framework 7 program in 2010.\\
The coordination part of this project is funded under the European project ESPAS 7th FP RI 283676.
\end{acknowledgements}

\end{document}